\documentclass[twocolumn,showpacs,aps,prd,superscriptaddress,amsmath,amssymb,floatfix]{revtex4}
\usepackage{graphicx}
\usepackage{amsmath}
\usepackage{bm}
\usepackage{color}

\begin{document}
\title{Decay of the Higgs boson to $\tau^- \tau^+$ and non-Hermiticity of the Yukawa interaction}
\author{Alexander Yu. Korchin} \email{korchin@kipt.kharkov.ua}
\affiliation{NSC `Kharkov Institute of Physics and Technology',
61108 Kharkiv, Ukraine}
\affiliation{V.N.~Karazin Kharkiv National University, 61022 Kharkiv, Ukraine}
\author{Vladimir A. Kovalchuk}  \email{koval@kipt.kharkov.ua}
\affiliation{NSC `Kharkov Institute of Physics and Technology',
61108 Kharkiv, Ukraine}

\date{today}

\begin{abstract}
The issue of Hermiticity of the Higgs boson interaction with fermions is addressed. A model for non-Hermitian Yukawa interaction is proposed and
approximation of one fermion generation is considered. Symmetry
properties of the corresponding $h f \bar{f}$ Lagrangian with
respect to the discrete $\mathcal{P}$, $\mathcal{C}$ and
$\mathcal{T}$ transformations are analyzed, and the modified
Dirac equation for the free fermion is studied.
Longitudinal polarization of the fermions in the decay $h \to f \bar{f}$, which arises due to non-Hermiticity of the $h f \bar{f}$ interaction, is discussed. It is suggested to study effects of this non-Hermiticity in the decay $h \to \tau^- \tau^+ \to \mu^- {\bar \nu}_\mu \nu_\tau \, \mu^+ \nu_\mu {\bar \nu}_\tau$, for which observables (asymmetries) are constructed which take nonzero values for a non-Hermitian $h \tau^- \tau^+$ interaction. These asymmetries are analyzed for various configurations of the muon energies.
\end{abstract}

\pacs{11.30.Er, 12.15.Ji, 12.60.Fr, 14.80.Bn}

\maketitle

\setcounter{footnote}{0}

\section{\label{sec:Introduction}Introduction}

In 2012 at the Large Hadron Collider (LHC) the Collaborations
ATLAS and CMS discovered the spinless particle $h$ with the mass
approximately equal to 125 GeV~ \cite{ATLAS:2012,CMS:2012}. The
study of the processes of $h$ boson production and decay modes
has shown that its properties are
consistent~\cite{CMS:2015,ATLAS:2016} with the properties of the
Higgs boson of the Standard model (SM). In particular, analysis of
the angular correlations in the $h\to
ZZ^*,\,Z\gamma^*,\,\gamma^*\gamma^*\to 4\ell$, $h\to WW^*\to
\ell\nu\ell\nu$ ($\ell=e,\,\mu$), and $h\to \gamma\gamma$ decay
modes has shown that all the data agree with the prediction for
the Higgs boson with the quantum numbers $J^{\rm
PC}=0^{++}$~\cite{CMS:2012sp,ATLAS:2013,CMS:2015sp}. Thus based on
these data one can conclude that the structure of the
$hWW$ and $hZZ$ interactions is in agreement with the SM.

In the SM the fermion masses are generated through the Yukawa
couplings between the Higgs field and the fermion fields.
Measurement of these couplings is needed for identification of
the particle $h$ with the SM Higgs boson.
At present, the intensity of the Higgs signal $\mu$, defined as the
ratio of the experimentally measured production cross section of
the Higgs boson with its subsequent decay to a set of final
particles $X$ to the corresponding value predicted in the SM, is
determined for the channels  $h\to \tau^-\tau^+$ and $h\to b\,
\bar b$. Namely, the ATLAS Collaboration obtained the values
$\mu(\tau^-\tau^+)=1.43^{+0.43}_{-0.37}$
\cite{ATLAS:2016,ATLAS:2015tau} and $\mu(b\bar b)=0.52 \pm
0.32\pm0.24$ \cite{ATLAS:2016,ATLAS:2015bb}, while the CMS
Collaboration obtained $\mu(\tau^-\tau^+)=0.91\pm 0.28$
\cite{CMS:2015}, $\mu(\tau^-\tau^+)=0.78\pm 0.27$
\cite{CMS:2014tau} and $\mu(b\bar b)=0.84 \pm 0.44$
\cite{CMS:2015}, $\mu(b\bar b)=1.0 \pm 0.5$ \cite{CMS:2014bb}.
Recently there appeared the combined ATLAS and CMS
measurements of the Higgs boson production and decay rates as
well as constraints on its couplings to vector bosons and fermions
\cite{ATLASCMS:2016}.  As a result the value of $\mu$ turns out to be equal to $1.09\pm0.11$.

The Lagrangian of the SM is invariant under the local transformations of
the group $SU(3)_C\otimes SU(2)_L\otimes U(1)_Y$ which is
spontaneously broken to the $SU(3)_C\otimes U(1)_{\rm QED}$ group.
This Lagrangian is built on the basis of the principle of minimal
coupling from Lagrangian of the free fermion fields and the scalar fields,
which is invariant under the global $SU(3)_C\otimes SU(2)_L\otimes U(1)_Y$
transformations.
The latter Lagrangian contains kinetic-energy terms
for the left- and right-chiral fermion fields and kinetic-energy
terms for the scalar fields, which are automatically Hermitian, and nontrivial self-interaction of the scalar fields, generating the spontaneous breaking of the electroweak symmetry, which is usually chosen Hermitian.

After replacing the derivative $\partial_\mu$  by the covariant
derivatives $D_\mu$, and adding gauge-invariant kinetic terms
for the gauge fields, one obtains the SM Lagrangian of the
massless fermions, which is Hermitian and symmetric under the local
$SU(3)_C\otimes SU(2)_L\otimes U(1)_Y$ transformations.

As for the Lagrangian describing the Yukawa interaction between
the fermion fields and the scalar fields, $\, \mathcal{L}_{\rm
Yuk}^{\rm SM}$, in the SM, in addition to the gauge invariance
under the  $SU(3)_C\otimes SU(2)_L\otimes U(1)_Y$ transformations,
the requirement of Hermiticity  of $\, \mathcal{L}_{\rm Yuk}^{\rm
SM}$ is imposed. Thus, unlike the other terms in the total SM
Lagrangian which are naturally Hermitian, the Yukawa interaction
has ``acquired'' Hermiticity which may not be necessary. In this
connection it seems important to verify whether the interaction of
the Higgs boson with fermions is Hermitian.

Note that models which are described by non-Hermitian Hamiltonians attracted interest for a long time~\cite{Bender:1998,Bender:2007}.
Recently in Ref.~\cite{Alexandre:2015} a non-Hermitian Yukawa
interaction between neutrino and scalar fields has been studied in
the SM and in its various extensions.

Some aspects of non-Hermiticity of the Higgs boson interaction with the
top quark have been addressed in
Refs.~\cite{Korchin:2013a,Korchin:2013b,Korchin:2014}. In
particular, in \cite{Korchin:2013a,Korchin:2013b} the polarization
characteristics of the photon in the decays $h \to \gamma \gamma$
and $h \to \gamma Z $ have been studied. The photon circular
polarization in these processes arises due to the
$\mathcal{CP}$-even and $\mathcal{CP}$-odd  components of the $h t
\bar{t}$ interaction, small imaginary loop contributions in the SM, and non-Hermiticity of the $h t \bar{t}$ \ interaction.
In \cite{Korchin:2014} it has been shown that the forward-backward
lepton asymmetry $A_{FB}$ in the processes $h \to \gamma \ell^+
\ell^-$ (for $\ell= e, \mu, \tau$) is sensitive to non-Hermiticity
of the Higgs interaction with the top quark, and $A_{FB}$ can
acquire values of about 15 \% (20 \%) for the muon-antimuon
(electron-positron) pairs.

We also emphasize  that measurement of
any observable sensitive to non-Hermiticity of the Lagrangian
can be used at the same time for testing the $\mathcal{CPT}$
theorem, since Hermiticity of the Hamiltonian (or Lagrangian) is a
necessary condition in the proof of the $\mathcal{CPT}$ theorem in quantum field theory (see, {\it e.g.}, \cite{Streater:1964}).
In this connection we recall Ref.~\cite{Okun:2002}, where the close relation of non-Hermiticity with violation of the $\mathcal{CPT}$ symmetry has been noted. The author of \cite{Okun:2002} has also shown that observation of photon circular polarization in the pion decay $\pi^0 \to \gamma \gamma$, or muon longitudinal polarization in the $\eta$-meson decay $\eta \to \mu^- \mu^+$,  would be a signal of violation of the  $\mathcal{CPT}$ symmetry.

In the present paper we study effects of non-Hermiticity of the
Yukawa interaction in the Higgs boson decay to pair of $\tau$
leptons. Note that though the decay $h\to \tau^- \tau^+$ has been
discussed in literature  (see, for instance, \cite{Djouadi:2008}
and references therein) for a long time, the main interest there
has been concentrated on investigation of effects of the
$\mathcal{CP}$--symmetry violation in the Hermitian Yukawa
interaction and differences in the angular distributions for the
scalar and pseudoscalar Higgs boson. In the present paper, in
contrast, we mainly pay attention to observables which are sensitive
to non-Hermiticity of the Yukawa interaction. In addition, a model
for non-Hermitian Yukawa interaction in case of one fermion
generation is proposed.

The paper is organized as follows. In Sec.~\ref{sec:formalism} the
decay width of the Higgs boson to the polarized fermion $f$ and
antifermion $\bar{f}$ is considered, and polarization characteristics of
$f  \, (\bar{f})$ are discussed. The fully differential
width of the decay into the lepton channel $h \to \tau^- \tau^+
\to \mu^- {\bar \nu}_\mu \nu_\tau \, \mu^+ \nu_\mu {\bar
\nu}_\tau$ is derived and the distribution over the muon energies
is obtained. Observables are proposed which carry information
on the  $h \tau^- \tau^+$ non-Hermiticity. In Sec.~\ref{sec:model}
a model for non-Hermitian Yukawa interaction between the Higgs
fields and fermions is proposed and the approximation of one
fermion generation is studied. In Sec.~\ref{sec:results} results
of calculation and discussion are presented. In
Sec.~\ref{sec:conclusions} we draw conclusions. In Appendix~A
functions $f(x_1,x_2)$ and $g(x_1,x_2)$, which enter the
distribution over the muon energies, are defined.


\section{\label{sec:formalism} Decays ${h \to f \bar{f}}$ and
${ h \to \tau^- \tau^+ \to \mu^-  \mu^+ + 4 \; {\rm neutrinos}}$ }


We assume that the couplings of $h$ boson to the fermion fields,
$\psi_f$, are given by the Lagrangian including both scalar and
pseudoscalar parts
\begin{equation}\label{eq:2001}
{\cal L}_{hff}=-\sum_{f = \ell, \, q} \frac{m_f}{v}\,h\,{\bar
\psi_f}\left(a_f+i\,b_f \gamma_5\right)\psi_f \,,
\end{equation}
where $v=\left(\sqrt{2}G_{\rm F}\right)^{-1/2}\approx 246$ GeV is
the vacuum expectation value of the Higgs field,  $G_F =
1.1663787(6) \times 10^{-5}$  GeV$^{-2}$ is the Fermi
constant~\cite{PDG:2014}, $m_f$ is the fermion mass and $a_f$,
$b_f$ are complex parameters
($a_f=1$ and $b_f=0$ corresponds to the SM).
At the same time, the Higgs interaction with the $W^\pm$ and $Z$
bosons is chosen as in the SM. In terms of these parameters the
decay width of the Higgs to unpolarized fermions, except the top quark, in the leading order is
equal to
\begin{equation}
\Gamma (h \to f \bar{f})\, = \, \frac{N_f G_F}{4 \sqrt{2} \pi} \,
m_f^2 \, m_h \, \beta_f \bigl( |a_f|^2 \beta_f^2 \, + \, |b_f|^2
\bigr) \,, \label{eq:2002}
\end{equation}
where $\beta_f = \sqrt{1- 4m_f^2/m_h^2} $ is the fermion velocity
in the rest frame of $h$, $N_f = 1 (3)$ for leptons (quarks).
Apparently one can put  $\beta_f \approx 1$.

For the real parameters $a_f$ and $b_f$ the interaction (\ref{eq:2001}) is Hermitian,
however it is seen from Eq.~(\ref{eq:2002})
that any non-Hermiticity of the Lagrangian Eq.~(\ref{eq:2001})
does not affect the width of the Higgs boson decay to fermions. In addition, if parameters $a_f, \, b_f$ either satisfy the equation
\begin{equation}
|a_f|^2 + |b_f|^2 = 1 \,, \label{eq:2003}
\end{equation}
or the expression $|a_f|^2 + |b_f|^2$ turns out close to unity, then the
$h \to f \bar{f}$ decay width will have the same value as in the SM, or close to it.

However, the situation changes if it will become possible to
measure the polarization characteristics of the fermions. Indeed,
the rate of the Higgs boson decay to polarized fermions is
determined by the expression
\begin{eqnarray}
\frac{d\Gamma}{d\Omega}&=&\Gamma (h \to f
\bar{f})\frac{1}{16\pi}\Bigl(1-\zeta_{1L}\zeta_{2L}+\frac{|a_f|^2
\beta_f^2 - |b_f|^2}{|a_f|^2 \beta_f^2 + |b_f|^2}\nonumber
\\ &\times&(\vec{\zeta}_{1T}\cdot\vec{\zeta}_{2T})-
\frac{2\,{\rm Re}(a_f\,b_f^*)}{|a_f|^2 \beta_f^2 +
|b_f|^2}\beta_f\,\vec{n}\cdot[\vec{\zeta}_{1T}\times\vec{\zeta}_{2T}]\nonumber
\\ &-&\frac{2\,{\rm Im}(a_f\,b_f^*)}{|a_f|^2 \beta_f^2 +
|b_f|^2}\beta_f\,(\zeta_{1L}-\zeta_{2L} )\Bigr), \label{eq:2004}
\end{eqnarray}
where $\vec{\zeta}_1$ ($\vec{\zeta}_2$) is the polarization vector of the fermion $f$ ($\bar f$) in the rest frame
of $f$ ($\bar f$),  $\vec{n}$ is the unit  vector in the
direction of 3-momentum of fermion $f$ in the rest frame of the
$h$ boson. Further, the longitudinal and transverse components of
polarization are defined as
$\zeta_{iL} \equiv (\vec{\zeta}_i \cdot \vec{n})$ and
$\vec{\zeta}_{iT}\equiv \vec{\zeta}_i - \vec{n} (\vec{\zeta}_i \cdot \vec{n})$,
where $i=1\,,2$. Note that the covariant form of the rate of the
Higgs boson decay to polarized fermions has been considered in
Refs.~\cite{Djouadi:2008,Arens:1994}.
In Ref.~\cite{Bernreuther:1997} the spin density matrix of the $f \bar{f}$ system has been calculated for the  decays $h \to t \bar{t} $ and $h \to \tau^+ \tau^-$ with account of radiative corrections of the order $\alpha_s$ and $\alpha_{\rm em}$, respectively.
The interaction (\ref{eq:2001}) with real parameters $a_f$ and $b_f$ has been used in \cite{Bernreuther:1997}.

We see that for the non-Hermitian Lagrangian in
Eq.~(\ref{eq:2001}) the fermion $f$ ($\bar f$) is longitudinally
polarized with polarization equal to
\begin{equation}
\alpha_L = \frac{2\,|{\rm Im}(a_f\,b_f^*)|}{|a_f|^2 \beta_f^2 +
|b_f|^2}\beta_f. \label{eq:2005}
\end{equation}
The direction of fermion polarization is opposite to the direction
of its movement (or in the direction of its movement) depending on
the sign of the quantity ${\rm Im}(a_f\,b_f^*)/|{\rm
Im}(a_f\,b_f^*)|=\pm 1$.

Note that presence of both parameter $a_f$ and $b_f$ in (\ref{eq:2004}), which leads to the $\mathcal{CP}$ violation in the Higgs boson interaction with fermions, manifests itself not only in the longitudinal polarization of the fermion but also in the nonzero spin-spin correlation term $ \propto {\rm Re}(a_f\,b_f^*) \; \vec{n} \cdot [\vec{\zeta}_{1T}\times\vec{\zeta}_{2T}]$.

Of course, measurement of the polarization of the final fermions in the decay $h \to f \bar{f}$ is a difficult problem.
Moreover, measurement of the polarization of the $b$- and $c$-quarks,
created on the LHC, is itself an important task independently from
their production mechanism.  In principle, as has been shown in
\cite{Galanti:2015}, the ATLAS and CMS can measure the
polarization of the $b$ quark by using the semileptonic decay of
$\Lambda_b$ baryon, and the polarization of the $c$ quark using
the decay of $\Lambda_c$ baryon, $\Lambda_c^+ \to p K^- \pi^+$, created in the QCD collisions and coming from the decay of the top quark.

In general, the longitudinal polarization (\ref{eq:2005}) of the fermion can also arise due to radiative corrections which generate imaginary part of the $h \to f \bar{f}$ amplitude.  Such corrections for the $t \bar{t}$ and $\tau^+ \tau^-$ pairs are calculated in \cite{Bernreuther:1997} with the Hermitian Lagrangian (\ref{eq:2001}) for real $a_f, \, b_f$. In particular, for the case of the $\tau$ leptons, the QED radiative corrections, or the $\tau^+ \tau^-$ rescattering via the photon exchange, are shown to give a negligibly small contribution
of the order $\alpha_{\rm em} (m_h) \times (m_\tau / m_h)^2 \approx 10^{-6}$
to the longitudinal polarization of the $\tau$ lepton.  Based on this observation the authors of
\cite{Bernreuther:1997} concluded that this polarization is not a useful tool for analyzing the ${\cal CP}$ nature of the Higgs boson.

In the SM, the other possible one-loop corrections to the $h \to \tau^+ \tau^-$ amplitude arise due to intermediate $W^+ W^-$-bosons, $ZZ$-bosons and neutrino $\nu_\tau \bar{\nu}_\tau$, however the former two contributions are real since $m_h < 2 m_W, \, 2m_Z$, and the latter one is extremely small and can be safely neglected.

In models beyond the SM, the imaginary part of one-loop diagrams could arise from some intermediate particles $X$ in the loops with the masses $m_X < m_h /2$. This would imply a possibility of the Higgs-boson decay $h \to X \bar{X}$, however no new particles beyond the SM have been observed at the LHC so far. In any case the QED radiative correction is probably  the dominant, but very small contribution to the longitudinal polarization of the $\tau$ lepton.
Therefore if the degree of this polarization turned out to be different from prediction of Ref.~\cite{Bernreuther:1997}, {\it e.g.} much larger, then it would point out to a non-Hermiticity of the $h \tau^+ \tau^-$ interaction.

Here we will not discuss the Higgs boson decay modes to quarks and consider
the decay of $h$ boson to $\tau^-\,\tau^+$ pair with their
consequent decay into the channels $\tau^-\to \mu^-{\bar \nu}_\mu
\nu_\tau$ and $\tau^+\to \mu^+\nu_\mu {\bar \nu}_\tau$.
The differential decay width of the decay $h(p)\to
\tau^-(k_1)+\tau^+(k_2)\to \mu^-(p_1){\bar \nu}_\mu \nu_\tau +
\mu^+(p_2)\nu_\mu {\bar \nu}_\tau$ is
\begin{widetext}
\begin{eqnarray}
d\Gamma&=&\Gamma(h\to\tau^-\tau^+)\left(\frac{\tau\,G_F^2}{48\pi^4}\right)^2
\frac{d^3\vec{p}_1}{E_1}\frac{d^3\vec{p}_2}{E_2}\biggl(s_1s_2(s_1+s_2)
-m^2((s_1+s_2)^2-y(s_1^2+s_2^2-s_1s_2))+m^4(1-y^2)(s_1+s_2)\nonumber
\\&-&m^6y(1-y)^2+\bigl(4s_1s_2-2m^2(1-y)
(s_1+s_2)+m^4(1-y)^2\bigr)\Bigl(\frac{|a|^2\beta^2-|b|^2}{|a|^2\beta^2+|b|^2}
(p_1\cdot p_2)+\frac{2|a|^2}{|a|^2\beta^2+|b|^2}((k_1-k_2)\cdot
p_1)\nonumber
\\&\times&((k_1-k_2)\cdot p_2)/m_h^2
+\frac{2|b|^2}{|a|^2\beta^2+|b|^2}(p\cdot p_1)(p\cdot
p_2)/m_h^2+\frac{4\,{\rm
Re}(ab^*)}{|a|^2\beta^2+|b|^2}\varepsilon_{\mu\nu\rho\sigma}p^\mu
k_1^\nu p_1^\rho p_2^\sigma/m_h^2\Bigr)+\frac{2\,{\rm
Im}(ab^*)}{|a|^2\beta^2+|b|^2}\Bigl((s_1\nonumber
\\&-&s_2)(s_1s_2-m^2(1-y)(s_1+s_2-m^2))+(m^6(1-y)^3-2m^2(1+y)s_1s_2) p\cdot (p_1-p_2)/m_h^2-2m^2(1-y)
(s_2^2p\cdot p_1\nonumber
\\&-&s_1^2p\cdot
p_2)/m_h^2+(4s_1s_2+m^4(1-y^2))(s_2p\cdot p_1-s_1 p\cdot
p_2)/m_h^2-2m^4(1-y)^2(s_1 p\cdot p_1-s_2 p\cdot p_2)/m_h^2
\Bigr)\biggr), \label{eq:2006}
\end{eqnarray}
\end{widetext}
where $p$, $k_1$ and $k_2$, $p_1$ and $p_2$ are the 4-momenta of
$h$ boson, $\tau^-$ and $\tau^+$ leptons,  $\mu^-$ and $\mu^+$
muons, respectively, $p_1=(E_1,\,\vec{p}_1)$,
$p_2=(E_2,\,\vec{p}_2)$, $m$ is the mass of $\tau^\pm$ lepton,
$y=m_\mu^2/m^2$, $m_\mu$ is the mass of the muon, $\tau$ is the
lifetime of the $\tau^\pm$ lepton. Further $s_1=(k_1-p_1)^2$,
$s_2=(k_2-p_2)^2$, $\varepsilon_{\mu\nu\rho\sigma}$ is Levi-Civita antisymmetric symbol with
$\varepsilon_{0123}=+1$, and $\beta$ is the
$\tau^\pm$-lepton velocity in the rest frame of $h$ boson. We also
introduced the shortened notation $a \equiv a_\tau$ and $b \equiv b_\tau$.

After integration of Eq.~(\ref{eq:2006}) over the polar and
azimuthal angles we obtain the decay width as a function of the
energies of muons
\begin{eqnarray}
\frac{d\Gamma}{dx_1dx_2}&=&\Gamma(h\to\tau^-\tau^+)\left(\tau\,\frac{G_F^2m^5}{192\pi^3}\right)^2
\frac{8a(x_1)a(x_2)}{\beta^2(1+\beta)^5}\nonumber
\\&\times&\Bigl(f(x_1,x_2)+f(x_2,x_1)\nonumber
\\&+&\frac{2\,{\rm
Im}(ab^*)}{|a|^2\beta^2+|b|^2}\bigl(g(x_1,x_2)-g(x_2,x_1)\bigr)\Bigr),
\label{eq:2007}
\end{eqnarray}
where $x_1\equiv2E_1/m_h$ and $x_2\equiv2E_2/m_h$ are the
fractions of the energies of $\mu^-$ and $\mu^+$, which vary
within the limits
\begin{eqnarray}
&& x_{\rm min} \leq x_{1(2)}\leq x_{\rm max}, \label{eq:2008} \\
&& x_{{\rm max} / {\rm min} } = \frac{1 \pm \beta}{2}+y\frac{1 \mp \beta}{2}.
\label{eq:2008a}
\end{eqnarray}
The functions $f(x_1, x_2)$, \ $g(x_1, x_2)$ and $a(x)$ are defined in Appendix~A.

It is seen from Eq.~(\ref{eq:2007}) that in any Hermitian model of the $h f \bar{f}$ interaction, in which ${\rm Im}(ab^*) =0$, the differential width (\ref{eq:2007}) has the same form as in the SM.

In connection with Eq.~(\ref{eq:2007}) we should mention Ref.~\cite{Arens:1994}, where a similar equation was obtained for the decay
$h \to t \, \bar{t} \to \ell^+ \, \ell^- + \ldots \, $ under assumption that the $h$-boson is sufficiently heavy (400 GeV) to decay into the on-mass-shell top quarks, and in the narrow-width approximation for the $W$-boson
\footnote{Despite a similarity of Eq.~(\ref{eq:2007}) with Eq.~(6) from \cite{Arens:1994} there are essential differences in the functions defining energy-symmetric and energy-asymmetric parts of these equations. In Ref.~\cite{Arens:1994} the decay $h \to t \, \bar{t} \to \ell^+ \, \ell^- + \ldots \, $ proceeds through the two sequential two-body decays of the top quark (and antiquark),
$h \to t \, \bar{t} \to W^+ \, b \, + \, W^- \, \bar{b} \to  \ell^+ \, \nu_\ell \, b + \ell^- \, \bar{\nu}_\ell \,  \bar{b} $ with the $W$-bosons on the mass shells.
In contrast, in derivation of Eq.~(\ref{eq:2007}) the three-body decay of the $\tau$-lepton is assumed,
{\it i.e.} $h \to \tau^+ \tau^- \to \mu^+ \, \nu_\mu \, \bar{\nu}_\tau \,  + \, \mu^- \, \bar{\nu}_\mu\,  \nu_\tau$. Different reaction mechanisms lead to different analytical results [{\it cf.}, Eqs.~(\ref{eq:A001}),
(\ref{eq:A002}) in Appendix \ref{sec:Appendix} with equations in \cite{Arens:1994} following Eq.~(8)]. }.

It is convenient in addition to the differential
decay width in Eq.~(\ref{eq:2007}) to define the distribution over
the fractions of the muon energies
\begin{eqnarray}
&& W(x_1, \, x_2) \equiv \frac{1}{\Gamma } \frac{d \Gamma}{dx_1 dx_2}, \label{eq:2008b} \\
&& \Gamma = \Gamma(h\to \tau^- \tau^+) \, \Bigl( {\rm BR} (\tau^-
\to \mu^- \nu_\tau \bar{\nu}_\mu ) \Bigr)^2. \nonumber
\end{eqnarray}
This distribution is normalized to unity
\[ \int_{x_{\rm min}}^{x_{\rm max}} dx_1  \int_{x_{\rm min}}^{x_{\rm max}} dx_2 \, W(x_1, \, x_2) =1, \]
where $x_{\rm min}$ and $x_{\rm max}$ are defined in
(\ref{eq:2008a}) and are equal respectively to 0.00373716  and
0.999799. Then the fraction of the total number of muons, which corresponds to $\mu^-$ in the energy interval $[\varepsilon_1,
\varepsilon_1^\prime] $ and $\mu^+$ in the energy interval $ [\varepsilon_2,\varepsilon_2^\prime]$, is
\begin{equation} N (\varepsilon_1, \varepsilon_1^\prime; \,
\varepsilon_2, \varepsilon_2^\prime) =
\int_{\varepsilon_1}^{\varepsilon_1^\prime} dx_1
\int_{\varepsilon_2}^{\varepsilon_2^\prime} dx_2 \, W(x_1, \,
x_2), \label{eq:2008c}
\end{equation}
where the integration limits satisfy the conditions
$x_{\rm min} \le \varepsilon_{1 (2)} \le \varepsilon^{\prime}_{1 (2)} \le x_{\rm max} $.

Now we construct observable proportional to ${\rm Im} (ab^*)$.
Let us define asymmetry in the following way
\begin{equation}{\mathcal A}(\varepsilon_1, \varepsilon_1^\prime;  \varepsilon_2,\varepsilon_2^\prime)
\equiv \dfrac{N(\varepsilon_1, \varepsilon_1^\prime; \,
\varepsilon_2, \varepsilon_2^\prime)-N(\varepsilon_2,
\varepsilon_2^\prime; \, \varepsilon_1,
\varepsilon_1^\prime)}{N(\varepsilon_1, \varepsilon_1^\prime; \,
\varepsilon_2, \varepsilon_2^\prime)+N(\varepsilon_2,
\varepsilon_2^\prime; \, \varepsilon_1,
\varepsilon_1^\prime)}. \label{eq:2009}
\end{equation}
Using expression (\ref{eq:2007}) one can write for the asymmetry
\begin{equation}
{\mathcal
A}(\varepsilon_1,\varepsilon_1^\prime;\varepsilon_2,\varepsilon_2^\prime)=\frac{2\,{\rm
Im}(ab^*)}{|a|^2\beta^2+|b|^2}\, \Delta
(\varepsilon_1,\varepsilon_1^\prime;\varepsilon_2,
\varepsilon_2^\prime),\label{eq:2010}
\end{equation}
where
\begin{eqnarray}
&&\Delta (\varepsilon_1,\varepsilon_1^\prime;\varepsilon_2,\varepsilon_2^\prime)  \nonumber
\\&&=\int_{\varepsilon_1}^{\varepsilon_1^\prime} \! \! dx_1 \!\! \int_{\varepsilon_2}^{\varepsilon_2^\prime} \!\! dx_2
a(x_1)a(x_2)\bigl(g(x_1,x_2)-g(x_2,x_1)\bigr)\label{eq:2011}
\\&&
\times \Biggl( \! \int_{\varepsilon_1}^{\varepsilon_1^\prime} \!\! dx_1 \!\! \int_{\varepsilon_2}^{\varepsilon_2^\prime} \!\!  dx_2
a(x_1)a(x_2)\bigl(f(x_1,x_2)+f(x_2,x_1)\bigr) \! \Biggr)^{-1}. \nonumber
\end{eqnarray}

The asymmetry (\ref{eq:2010}) is nonzero for a non-Hermitian $h \tau^- \tau^+$ interaction. Its value is determined by the parameters $a$ and $b$
through ${\rm Im}(ab^*)$, and also essentially depends on the choice of the area $[\varepsilon_1, \varepsilon_1^\prime] \otimes [\varepsilon_2,\varepsilon_2^\prime]\cup[\varepsilon_2,
\varepsilon_2^\prime] \otimes [\varepsilon_1,\varepsilon_1^\prime]$ in which the energies of $\mu^-$ and $\mu^+$ vary in Eq.~(\ref{eq:2011}).

Along with the asymmetry (\ref{eq:2009}) and (\ref{eq:2010}) we can define the asymmetry of the $\mu^-$ and $\mu^+$ mean energies, namely
\begin{equation}
{\mathcal A}_{\rm E}\equiv\frac{\langle E_1\rangle-\langle
E_2\rangle}{\langle E_1\rangle+\langle E_2\rangle},
\label{eq:2011b}
\end{equation}
which is also proportional to ${\rm Im} (ab^*)$. Indeed, using Eq.~(\ref{eq:2007}) one can write
\begin{equation}
{\mathcal A}_{\rm E}=\frac{2\,{\rm
Im}(ab^*)}{|a|^2\beta^2+|b|^2}\: \delta_{\rm E},\label{eq:2012}
\end{equation}
where
\begin{eqnarray}
\delta_{\rm E} &=&\int_{x_{\rm min}}^{x_{\rm max}}\! \! \! x_1 dx_1 \!\!
\int_{x_{\rm min}}^{x_{\rm max}} \!\! dx_2
\, a(x_1)a(x_2)\Bigl(g(x_1,x_2)\nonumber \\
&-&g(x_2,x_1)\Bigr) \Biggl( \! \int_{x_{\rm min}}^{x_{\rm max}}
\! \! \! x_1 dx_1 \!\! \int_{x_{\rm min}}^{x_{\rm max}}
\!\!  dx_2 \, a(x_1)a(x_2)\nonumber  \\
&\times&\Bigl(f(x_1,x_2)+f(x_2,x_1)\Bigr) \! \Biggr)^{-1} .
\label{eq:2013}
\end{eqnarray}

From the definitions (\ref{eq:2011}) and (\ref{eq:2013}) it follows that $\Delta (\varepsilon_1,\varepsilon_1^\prime;\varepsilon_2,\varepsilon_2^\prime) $ and $\delta_{\rm E}$ can be calculated independently of
the parameters $a$ and $b$. In Sec.~\ref{sec:results} we present results of their calculation.


\section{\label{sec:model} A model for non-Hermitian Yukawa interaction  }

In the SM the Yukawa interaction Lagrangian of the Higgs field
with fermions satisfies the conditions of the gauge invariance and
Hermiticity. It has the form
\begin{eqnarray}
\mathcal{L}_{\rm Yuk}^{\rm
SM}&=&-\sum_{n\,,k}^3 \Bigl(f_{nk}^{(u)} \bar
q^{(0)}_{n L}\tilde{H}u^{(0)}_{kR}+f_{nk}^{(d)} \bar q^{(0)}_{n
L}H d^{(0)}_{kR}\nonumber \\&+&f_{nk}^{(e)} \bar \ell^{(0)}_{n L}H
e^{(0)}_{kR}+f_{nk}^{(\nu)} \bar \ell^{(0)}_{n L}\tilde{H}
\nu^{(0)}_{kR}\Bigr)+{\rm H.c.} \label{eq:3001}
\end{eqnarray}
In (\ref{eq:3001}) $n$ and $k$ are the generation indexes, and $L(R)$ refer to the left (right) chiral projections $\psi_{L(R)} \equiv \frac{1}{2}(1 \mp
\gamma_5)\psi$. The left-handed quarks and leptons
\begin{equation}\label{eq:3002}
q^{(0)}_{nL}=\left(\begin{array}{c} u^{(0)}_{nL}\\
d^{(0)}_{nL} \end{array} \right)\,,\quad
\ell^{(0)}_{nL}=\left(\begin{array}{c} \nu^{(0)}_{nL}\\
e^{(0)}_{nL} \end{array} \right)
\end{equation}
transform as $SU(2)$ doublets, while the right-handed fields
$u^{(0)}_{nR}$, $d^{(0)}_{nR}$, $\nu^{(0)}_{nR}$, and
$e^{(0)}_{nR}$  are singlets, in the weak-eigenstate
basis. In (\ref{eq:3001}) the matrices $f_{nk}$ describe the
Yukawa couplings between the single Higgs doublet $H$,
$\tilde{H}\equiv i\tau_2 H^*$, and the various flavors $n,\, k$
of quarks and leptons.

Now we omit the additional requirement of Hermiticity imposed on
the Yukawa interaction (\ref{eq:3001}) and choose
$\mathcal{L}_{\rm Yuk}$ in the form
\begin{eqnarray}
\mathcal{L}_{\rm
Yuk}&=&-\sum_{n\,,k}^3 \Bigl(f_{1nk}^{(u)} \bar
q^{(0)}_{n L}\tilde{H}u^{(0)}_{kR}+f_{1nk}^{(d)} \bar q^{(0)}_{n
L}H d^{(0)}_{kR}\nonumber \\&+&f_{1nk}^{(e)} \bar \ell^{(0)}_{n
L}H e^{(0)}_{kR}+f_{1nk}^{(\nu)} \bar \ell^{(0)}_{n L}\tilde{H}
\nu^{(0)}_{kR}\Bigr)\nonumber
\\&-&\sum_{n\,,k}^3 \Bigl(f_{2nk}^{(u)} \bar u^{(0)}_{nR}\tilde{H}^\dag
q^{(0)}_{k L}+f_{2nk}^{(d)} \bar d^{(0)}_{nR}H^\dag q^{(0)}_{k L}
\nonumber
\\&+&f_{2nk}^{(e)} \bar e^{(0)}_{nR}H^\dag\ell^{(0)}_{k L}
+f_{2nk}^{(\nu)} \bar \nu^{(0)}_{nR}\tilde{H}^\dag \ell^{(0)}_{k
L} \Bigr).\label{eq:3003}
\end{eqnarray}
It follows from (\ref{eq:3003}) that if $f_{2nk}\neq f_{1kn}^*$,
then the Yukawa interaction of the Higgs field with fermions does
not satisfy the Hermiticity requirement.

On this stage we will not study the consequences of the
non-Hermiticity of the Lagrangian (\ref{eq:3003}) on the flavor mixing.
We restrict ourselves to one generation and moreover take one
fermion from this generation. In this approximation the
Lagrangian describing the mass of the fermion, kinetic energy and
its interaction with the Higgs field $h$ can be presented in the
form
\begin{eqnarray}
{\cal L}(x)&=&-(1 + \frac{h(x)}{v}) \left( m_1
\bar{\psi}(x)\psi(x) + m_2
\bar{\psi}(x)\gamma_5\psi(x)\right)\nonumber \\
&+&\frac{i}{2}\left(\bar{\psi}(x)\gamma^\mu\partial_\mu \psi(x)-
\partial_\mu \bar{\psi}(x)\gamma^\mu\psi(x)\right),\label{eq:3004}
\end{eqnarray}
where $\psi(x)$ is the field of a fermion,
\begin{equation}
m_1=v\displaystyle\frac{f_1+f_2}{2\sqrt{2}}, \qquad
m_2=v\displaystyle\frac{f_1-f_2}{2\sqrt{2}},
\label{eq:3004_2}
\end{equation}
and $f_1$, $f_2$ are the Yukawa coupling constants. Note that
description of neutrino with non-Hermitian Yukawa interaction
has been studied in Ref.~\cite{Alexandre:2015}.

From the Lagrangian (\ref{eq:3004}) we obtain the modified Dirac
equation for the free fermion field
\begin{equation}
\label{eq:3005}
i  \gamma^\mu \frac{\partial\,\psi(x)}{\partial x^\mu} -
\left(m_1+ m_2 \gamma_5\right) \psi(x) = 0.
\end{equation}
This is the first-order differential equation. Acting by the
operator $i\gamma_\nu\partial^\nu$ on Eq.~(\ref{eq:3005}) we obtain
\begin{equation}\label{eq:3006}
\bigl(g_{\mu\nu}\partial^\mu\partial^\nu+m^2\bigr) \psi(x) = 0,
\end{equation}
where
\begin{equation}\label{eq:30061}
m^2=m_1^2-m_2^2 \quad {\rm or}\quad
m^2=\displaystyle\frac{v^2}{2}f_1f_2.
\end{equation}
Therefore, if $\psi(x)$ satisfies the Eq.~(\ref{eq:3005})
then each of the components of $\psi(x)$ has to obey the
Klein-Gordon equation (\ref{eq:3006}). It is clear that $m$ is the
mass of a fermion.

In the SM $f_2 = f_1^*$, so that $m^2=\displaystyle\frac{v^2}{2} |f_1|^2$
is the real-valued (in fact, in the SM $f_1$ can be made real and positive, {\it i.e.} $f_1 = f_2 \geq 0$).
While for non-Hermitian interaction, $m^2$ can be
real or complex. We note that the unstable particles are usually
characterized by the complex mass
\begin{equation}\label{eq:30062}
m^2=M^2-iM\Gamma , \quad {\rm or} \quad
m^2=\left(M-i\dfrac{\Gamma}{2}\right)^2,
\end{equation}
where $M$ and $\Gamma$ are their mass and width, respectively, while
the stable particles are characterized by the real mass. Of
course, the question on whether the interaction coupling of the
Higgs boson with fermions is real or complex requires experimental
study. At the same time, the experimental data on the total decay
width of the fundamental fermions~ \cite{PDG:2014} show that the
charged leptons, muon and $\tau$ lepton, have the width much
smaller than their mass. One has $\Gamma_\mu=2.84\times 10^{-18} \,
m_\mu$ and $\Gamma_\tau=1.28\times 10^{-12} \, m_\tau$, and the
electron can be considered as the stable particle. For neutrino
there exists only a constraint on the ratio of the mean
lifetime and the mass $\tau_\nu /m_\nu$, from which it follows
that if neutrino mass is not extremely small, then the width is
much smaller than the mass. Regarding the quarks, there is no
information on their decay width aside from the $t$ quark for which
$\Gamma_t=1.15\times 10^{-2} \, m_t$ with $m_t=173.21$ GeV.
Therefore, if the $h f \bar{f}$ coupling constant is proportional
to a complex mass coming from instability of a fermion, then its
influence on processes with participation of the Higgs boson and
fermions will probably be negligible, except for the top
quark.

These two possibilities, namely the real and complex parameters
$m_1$ and $m_2$, lead to drastically different behavior of the
Lagrangian density (\ref{eq:3004}) under the $\mathcal{P}$,
$\mathcal{C}$  and $\mathcal{T}$ transformations.
Under the space-inversion transformation, charge conjugation and
time inversion, $\psi(t\,,\vec{x})$ and $h(t\,,\vec{x})$ transform
as follows~\cite{Bigi:2000}
\begin{eqnarray}
&&\psi^P(t,\vec{x})=\gamma^0\psi(t,-\vec{x}),  \quad
h^P(t,\vec{x})=h(t,-\vec{x}),
\label{eq:3009}\\
&&\psi^C(x)=i \gamma^2\psi^*(x), \quad h^C(x)=h(x), \label{eq:3010}\\
&&\psi^T(t,\vec{x})= \gamma^1\gamma^3\psi(-t,\vec{x}), \; \,
h^T(t,\vec{x})= h(-t,\vec{x}),
\label{eq:3011}
\end{eqnarray}
respectively,  $\gamma^\mu$ are the matrices in the Pauli-Dirac
representation. Of course the second (kinetic-energy) term in (\ref{eq:3004}) is invariant with respect to $\mathcal{P}$, $\mathcal{C}$ and
$\mathcal{T}$ transformations, while the first term in
(\ref{eq:3004}), as will be seen below, is invariant with respect
to $\mathcal{C}$ transformation and not invariant under
$\mathcal{P}$ transformation for both real and complex values of
parameters $m_1$ and $m_2$. Regarding $\mathcal{T}$
transformation, the first term in (\ref{eq:3004}) is invariant for
real $m_1$ and $m_2$ and not invariant for complex parameters.
Indeed, using the definitions (\ref{eq:3009})--(\ref{eq:3011}) one
obtains:
\begin{eqnarray}
&&\mathcal{P}\left(1 +
\frac{h(t,\vec{x})}{v}\right)\overline{\psi}(t,\vec{x})\left(m_1+m_2\gamma_5\right)
\psi(t,\vec{x})\mathcal{P}^{-1}\nonumber
\\
&&=\left(1 +
\frac{h(t,-\vec{x})}{v}\right)\overline{\psi}(t,-\vec{x})\left(m_1-m_2\gamma_5\right)
\psi(t,-\vec{x}),\nonumber
\\
&&\mathcal{C}\left(1 +
\frac{h(x)}{v}\right)\overline{\psi}(x)\left(m_1+m_2\gamma_5\right)
\psi(x)\mathcal{C}^{-1}\nonumber
\\&&=
\left(1 +
\frac{h(x)}{v}\right)\overline{\psi}(x)\left(m_1+m_2\gamma_5\right)\psi(x), \label{eq:3012}\\
&&\mathcal{T}\left(1 +
\frac{h(t,\vec{x})}{v}\right)\overline{\psi}(t,\vec{x})\left(m_1+m_2\gamma_5\right)\psi(t,\vec{x})
\mathcal{T}^{-1}\nonumber
\\&&=
\left(1 +
\frac{h(-t,\vec{x})}{v}\right)\overline{\psi}(-t,\vec{x})\left(m_1^*+m_2^*\gamma_5\right)\psi(-t,\vec{x}).\nonumber
\end{eqnarray}
Thus, for real $m_1$ and $m_2$ the Lagrangian density
(\ref{eq:3004}) is not invariant under $\mathcal{P}$,
$\mathcal{CP}$, $\mathcal{PT}$ and $\mathcal{CPT}$ transformations
\footnote{Strictly speaking the symmetry arguments apply to the
corresponding action ${\cal S}=\int {\cal L}(x) \, d^4 x$.}. While
for complex $m_1$ and $m_2$ the Lagrangian density (\ref{eq:3004})
is not invariant under $\mathcal{P}$, $\mathcal{T}$,
$\mathcal{CP}$, $\mathcal{PT}$, $\mathcal{CT}$ and $\mathcal{CPT}$
transformations. These properties are summarized in
Table~\ref{tab:symmetries}.
Note that at the same time the Higgs boson interaction
with the $W^\pm$ and $Z^0$ bosons is $\mathcal{C}$,
$\mathcal{P}$ and $\mathcal{T}$ invariant.

\begin{table}[th]
\caption{Behavior of the Lagrangian (\ref{eq:3004}) under discrete
symmetries and Hermiticity. ``Yes''  (``No'') means that
the Lagrangian satisfies (does not satisfy) the symmetry.  }
\begin{center}
\begin{tabular}{c c c c }
\hline \hline
          & $m_1$--real,   & $m_1$--real,        & $m_1$--complex, \\
                    &  $m_2$--real   & $m_2$--imaginary & $m_2$--complex \\
\hline
$\mathcal{P}$   & No  & No  & No  \\
$\mathcal{C}$   & Yes & Yes & Yes \\
$\mathcal{T}$   & Yes & No  & No  \\
$\mathcal{CP}$  & No  & No  & No  \\
$\mathcal{PT}$  & No  & Yes & No  \\
$\mathcal{CT}$  & Yes & No  & No  \\
$\mathcal{CPT}$ & No  & Yes & No  \\
Hermiticity     & No  & Yes & No  \\
\hline \hline
\end{tabular}
\end{center}
\label{tab:symmetries}
\end{table}

Now we consider the case of real and positive constants $f_1$ and $f_2$.
Then the modified Dirac equation for the free fermion
(\ref{eq:3005}) and the Lagrangian density (\ref{eq:3004}) can be written as
\begin{equation}\label{eq:3007}
i\gamma^\mu \frac{\partial\,\psi(x)}{\partial x^\mu} -m\,e^{\xi
\gamma_5}\psi(x) = 0.
\end{equation}
and
\begin{eqnarray}
{\cal L}(x)&=&\frac{i}{2}\left(\bar{\psi}(x)\gamma^\mu\partial_\mu
\psi(x)-
\partial_\mu \bar{\psi}(x)\gamma^\mu\psi(x)\right)\nonumber \\
&-&m(1 + \frac{h(x)}{v})\bar{\psi}(x)e^{\xi \gamma_5}\psi(x),
\label{eq:3008}
\end{eqnarray}
where
\[\cosh\xi=\frac{m_1}{m},\quad \sinh\xi=\frac{m_2}{m}, \quad
m=v\sqrt{\frac{f_1f_2}{2}}.\]
Note that the Dirac equation with the fermion mass term in the form
$m_1+m_2\gamma_5$ has also been considered in Refs.~\cite{Bender:2005,Alexandre:2015a}.

From Eq.~(\ref{eq:3007}) one finds the positive energy,
$\psi^{(+)}(x) = \exp{(-i p \cdot x)} u(p)$ and the negative
energy, $\psi^{(-)}(x) = \exp{(+ i p \cdot x)} v(p)$, solutions.
The four-momentum and the energy of the fermion are
\begin{equation}\label{eq:3013}
p^\mu = (E_{p}, \vec{p}\,),  \qquad E_{p} = \left( \vec{p}^{\,2} +
m^2 \right)^{1/2} .
\end{equation}
In momentum space the modified Dirac equations for
the free fermion are
\begin{equation}\label{eq:3014}
\left(\not{\!p}-m e^{ \xi\gamma_5}\right)u_r(\vec{p}\,)=0,\quad
\left(\not{\!p}+m e^{\xi\gamma_5}\right)v_r(\vec{p}\,)=0,
\end{equation}
where $\not{\!p} \equiv \gamma^\mu p_\mu$ and $r=1,2$ labels
two independent solutions. They satisfy the following normalization conditions
\begin{equation}
\bar u_r(\vec{p}\,)
u_{r^\prime}(\vec{p}\,)=2m\,\delta_{r\,r^\prime},\; \; \bar
v_r(\vec{p}\,) v_{r^\prime}(\vec{p}\,)=-2m\,\delta_{r\,r^\prime}.
\label{eq:3015}
\end{equation}
The projection operators on the states with definite
polarization along the space-like 4-vector $s$ ($s^2=-1$),
orthogonal to $p$ ($s\cdot p=0$), are
\begin{eqnarray}
u(p,\,s)\bar u(p,\,s)&=&e^{- \xi/2\gamma_5}\left(\not{\!p}+
m\right)\frac{1+\gamma_5\not{\!s}}{2}e^{ \xi/2\gamma_5},\nonumber
\\ v(p,\,s)\bar
v(p,\,s)&=&e^{- \xi/2\gamma_5}\left(\not{\!p}-
m\right)\frac{1+\gamma_5\not{\!s}}{2}e^{ \xi/2\gamma_5}. \nonumber
\end{eqnarray}
The propagator of the free fermion has the form
\begin{eqnarray}
&&\langle 0|T\psi(x)\bar\psi(y)|0\rangle =iS_F(x-y),\nonumber \\
&&S_F(x-y)=\int
\frac{d^4p}{(2\pi)^4}e^{-ip\cdot(x-y)}\frac{\not{\!p}+m\,e^{-
\xi\gamma_5}}{p^2-m^2+i\epsilon}. \label{eq:3016}
\end{eqnarray}

In the Lagrangian model (\ref{eq:3008}) the rate of decay of the
Higgs boson to the polarized $\tau^-\,\tau^+$ leptons,
approximating the $\tau^\pm$ velocity in the $h$ rest frame by
unity, is determined from the expression
(\ref{eq:2004}) with  $a=\cosh\xi$ and $b=-i\sinh\xi$:
\begin{eqnarray}
\frac{d\Gamma}{d\Omega}&=&\Gamma (h \to
\tau^-\tau^+)\frac{1}{16\pi}\Bigl(1-\zeta_{1L}\zeta_{2L}+
\frac{\vec{\zeta}_{1T}\cdot\vec{\zeta}_{2T}}{\cosh2\xi}\nonumber
\\ &-&\tanh2\xi(\zeta_{1L}-\zeta_{2L} )\Bigr), \label{eq:3017}
\end{eqnarray}
where
\begin{equation}
\Gamma (h \to \tau^-\tau^+) = \frac{G_F\,m^2}{4 \sqrt{2}\, \pi} \,
m_h \cosh2\xi. \label{eq:3018}
\end{equation}
As $m^2=m_1^2-m_2^2$, in order to estimate the decay width $h \to
\tau^-\tau^+$ one has to know $m_2^2$. If $m_2^2$ comes from the
mean lifetime of the $\tau^\pm$ lepton, then
$m_2^2= \Gamma_\tau^2 / 4$. In this case $\xi\approx0$
and therefore the width of the $h \to \tau^-\tau^+$ decay practically
coincides with the width in the SM.

If $m_2^2$ has other origin, then the decay width of $h \to
\tau^-\tau^+$ can differ from the SM prediction. Indeed, let us
write the ratio of the $h\to \tau^-\tau^+$ decay width in the
model (\ref{eq:3008}) and in the SM, and the longitudinal polarization of the $\tau$ lepton,
\begin{eqnarray}
\kappa^2_\tau &\equiv& \frac{\Gamma (h \to \tau^- \tau^+)}{\Gamma_{\rm SM} (h
\to \tau^- \tau^+ )} \nonumber \\
& = & v^2\frac{f_1^2+f_2^2}{4 m^2 } = \frac{f_1^2+f_2^2}{2 f_1 f_2} \ge 1, \label{eq:3019} \\
\alpha_L &=& \frac{f_1^2-f_2^2}{f_1^2+f_2^2} .
\label{eq:3020}
\end{eqnarray}

In the SM $f_1=f_2 \equiv f_{\rm SM}=\sqrt{2}{m}/{v}$. Taking into account the constraint $m^2= v^2 f_1 f_2 /2$ we have
\begin{eqnarray}
f_1 & = & f_{\rm SM} \, e^{\xi}, \qquad f_2 = f_{\rm SM} \, e^{-\xi},   \label{eq:3021} \\
\kappa^2_{\tau} &=& \cosh 2\xi ,  \qquad  \alpha_L = \tanh2\xi ,
\label{eq:3022}
\end{eqnarray}
where $\xi=0$ corresponds to the SM and $f_{\rm SM}= 0.0102$ for the
$\tau$ lepton with mass 1.77682 GeV \cite{PDG:2014}.

\begin{figure}[tbh]
\begin{center}
\includegraphics[width=0.45\textwidth]{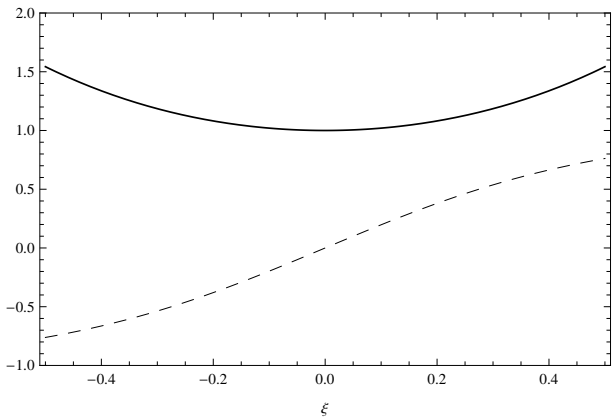}
\end{center}
\caption{Ratio $\kappa^2_\tau$ (solid line) and longitudinal
polarization $\alpha_L$ (dashed line) as functions of $\xi$.}
\label{fig:ratios}
\end{figure}

In Fig.~\ref{fig:ratios} the dependence of the ratio
(\ref{eq:3019}) and longitudinal polarization
(\ref{eq:3020}) on the parameter $\xi$ is presented. For an
estimate we choose the interval $ -0.5 \le \xi \le +0.5$.

As it is seen, the longitudinal polarization of the $\tau$ takes sizable
values, while the decay width varies not so much, up to a factor
of 1.5 for the ratio $\kappa^2_\tau $. Thus the values of the
measured $h\to \tau^-\tau^+$ decay width which are close to the
value in the SM will not necessarily mean that the structure of the
Yukawa interaction is the same as in the SM. Measurement of the $\tau$
longitudinal polarization is very important for obtaining
information on Hermiticity of the $h \tau^- \tau^+ $ interaction.


\section{\label{sec:results} Results of calculation and discussion}


In Table \ref{tab:function delta} we present results of
calculation  of the function
$\Delta(\varepsilon_1,\varepsilon_1^\prime; \,
\varepsilon_2,\varepsilon_2^\prime)$ in (\ref{eq:2011}) which
along with the factor $2\,{\rm Im}(ab^*)/
(|a|^2\beta^2+|b|^2)$ determines the asymmetry (\ref{eq:2010}).
It is seen that for certain intervals of the muon energies,
$\Delta(\varepsilon_1,\varepsilon_1^\prime; \,
\varepsilon_2,\varepsilon_2^\prime)$ takes quite big values.

\begin{table}[tbh]
\caption{Values of $\Delta(\varepsilon_1,\varepsilon_1^\prime; \, \varepsilon_2,\varepsilon_2^\prime)$ in (\ref{eq:2011}). The  intervals $[\varepsilon_1, \, \varepsilon_1^\prime]$ are indicated in the top raw, and the intervals $[\varepsilon_2, \, \varepsilon_2^\prime]$ -- in the left column.
 Note that $x_{\rm min} \le \varepsilon_{1 (2)} \le \varepsilon^{\prime}_{1 (2)} \le x_{\rm max}$.    }
\begin{center}
\begin{tabular}{c c c c c}
\hline \hline
  &  $[0.1, \, 0.3]$ & $[0.3, \, 0.5]$  & $[0.5, \,  0.7]$  &  $[0.7, 0.9]$ \\
\hline
$[0.1, \, 0.3]$    & 0.0   & -0.129   &  -0.327   & -0.593 \\
$[0.3, \, 0.5]$ & 0.129  &0.0  & -0.207  & -0.503 \\
 $[0.5, \,  0.7]$  & 0.327  & 0.207  & 0.0  & -0.330 \\
$[0.7, \, 0.9]$ & 0.593   & 0.503   &  0.330   & 0.0 \\
\hline \hline
\end{tabular}
\end{center}
\label{tab:function delta}
\end{table}
\begin{table}[tbh]
\caption{Fraction of the muon number $N(\varepsilon_1,\varepsilon_1^\prime; \, \varepsilon_2,\varepsilon_2^\prime)$ in Eq.~(\ref{eq:2008c}) in the SM. The  intervals $[\varepsilon_1, \, \varepsilon_1^\prime]$ are indicated in the top raw, and the intervals $[\varepsilon_2, \, \varepsilon_2^\prime]$ -- in the left column.
In the whole area $N(0.1,0.9; \, 0.1,0.9) = 0.7$.  }
\begin{center}
\begin{tabular}{c c c c c}
\hline \hline
     &  $[0.1, \, 0.3]$ & $[0.3, \, 0.5]$  & $[0.5, \,  0.7]$  &  $[0.7, 0.9]$ \\
\hline
$[0.1, \, 0.3]$    & 0.096  & 0.081 & 0.057 & 0.029
 \\
$[0.3, \, 0.5]$ &0.081 &    0.066   & 0.046 & 0.023
 \\
 $[0.5, \,  0.7]$  & 0.057  & 0.046 & 0.030 & 0.014
 \\
$[0.7, \, 0.9]$ &0.029 &    0.023    & 0.014 &   0.006
 \\
\hline \hline
\end{tabular}
\end{center}
\label{tab:probability}
\end{table}

One should keep in mind that feasibility of measuring the asymmetry
will  depend not only on values of
$\Delta(\varepsilon_1,\varepsilon_1^\prime; \,
\varepsilon_2,\varepsilon_2^\prime)$ and parameters $a, \, b$, but
also on the number of muons (\ref{eq:2008c}) in
this energy area. This fraction of the muon number is $ N(\varepsilon_1,
\varepsilon_1^\prime; \, \varepsilon_2, \varepsilon_2^\prime) + N(\varepsilon_2,
\varepsilon_2^\prime; \, \varepsilon_1, \varepsilon_1^\prime)$, and this number is independent of parameters $a, \, b$ and coincides with corresponding number calculated in the SM. We calculate $N(\varepsilon_1,
\varepsilon_1^\prime; \, \varepsilon_2, \varepsilon_2^\prime)$ in Table~\ref{tab:probability}. For any Hermitian interaction, the
function $N(\varepsilon_1, \varepsilon_1^\prime; \, \varepsilon_2, \varepsilon_2^\prime)$ is symmetric under the transformation $\varepsilon_1,
\varepsilon_1^\prime \leftrightarrow \varepsilon_2, \,
\varepsilon_2^\prime $.

Our analysis shows that the configuration of
the muons with energies close to the minimal allowed energy
$E_{1(2), {\rm min}} = x_{\rm min} \, {m_h}/{2} \approx 234$ MeV
is the most probable. In general, the smaller energies the muons have,
the bigger number of muons is.  This tendency is seen from Table~\ref{tab:probability}.
From Table~\ref{tab:function delta} it follows that in order to have big values of
$\Delta(\varepsilon_1,\varepsilon_1^\prime; \, \varepsilon_2,\varepsilon_2^\prime)$
one needs to choose $\mu^-$ and $\mu^+$ with big difference in energies.
Based on these observations we can take, for example,
\begin{equation}
x_1 \in [0.1, \, 0.3], \qquad x_2 \in [0.5, \, 0.7]
\label{eq:4003}
\end{equation}
with $\Delta(0.1, 0.3; \, 0.5,0.7) = 0.327$ and corresponding fraction of the number of muons  $N(0.1, 0.3; \, 0.5, 0.7) + N(0.1, 0.3; \, 0.5, 0.7) = 0.114$.

\begin{figure}[tbh]
\begin{center}
\includegraphics[width=0.45\textwidth]{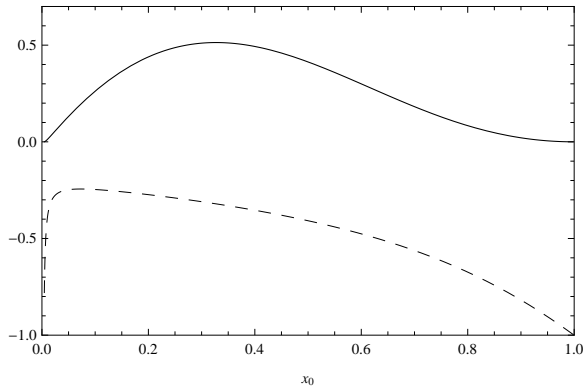}
\end{center}
\caption{Fraction of the number of muons $N(x_0)$ (solid line) and function $\Delta(x_0)$  (dashed line) vs. $x_0$. }
\label{fig:delta}
\end{figure}

In order to search for favorable conditions for the asymmetry we
consider the  following configuration of the muon energies.
Introduce an arbitrary ${x_0}$, such that $ x_{\rm min} \le x_0 \le
x_{\rm max}$, and calculate the function $\Delta(x_0)  \equiv \Delta(x_{\rm min}, {x_0}; \, {x_0}, x_{\rm max})$  and the fraction of the muon number
$N(x_0) \equiv N(x_{\rm min}, x_0 ; \, x_0,
x_{\rm max}) + N(x_0, x_{\rm max}; \, x_{\rm min}, x_0) $ for various values of $x_0$. Results of the calculation are presented in Fig.~\ref{fig:delta}.

It is seen from Fig.~\ref{fig:delta} that the
function $\Delta(x_0)$ reaches the value --1 at the ends of
the interval, i.e. at  $x_0 \approx x_{\rm min} =0.00373716$ and $x_0
\approx x_{\rm max} = 0.999799$. However the probability of these
configurations of the muons is close to zero. To have sizable values of $\Delta(x_0)$ and number of muons we can choose, for example,
\begin{equation}
x_0 \approx 0.6 ,  \,  \qquad |\Delta (x_0)|  \approx 0.5 , \qquad N(x_0) \approx 0.3.
\label{eq:4005}
\end{equation}
This means that muons should be selected in the intervals of energies
\begin{equation}
E_{\rm min} \, < \, E_{1 (2)}\, < \, E_0, \qquad  E_0 \, < \, E_{2 (1)}\,  <  \, E_{\rm max},
\label{eq:4006}
\end{equation}
where $E_{\rm min}= 234$ MeV,  \ $E_0 \approx  37.5$ GeV and $E_{\rm max} = 62.53$ GeV.

As for the asymmetry of mean muon energies (\ref{eq:2011b}) and (\ref{eq:2012}), direct calculation of coefficient $\delta_{\rm E}$ in (\ref{eq:2013}) gives
\begin{equation}
\delta_{\rm E} \approx 0.142.
\label{eq:4007}
\end{equation}

One can also study asymmetries of the $k$th moments of the energy
distribution (\ref{eq:2008b})
\begin{equation}
{\mathcal A}_{\rm E^k}\equiv\frac{\langle E_1^k\rangle-\langle
E_2^k\rangle}{\langle E_1^k\rangle+\langle E_2^k\rangle} =
\frac{2\,{\rm
Im}(ab^*)}{|a|^2\beta^2+|b|^2}\: \delta_{\rm E^k},
\label{eq:4008}
\end{equation}
with $\delta_{\rm E^2} \approx 0.249$, \ $\delta_{\rm E^3} \approx 0.332$, $\ldots$, which are more sensitive to the high-energy components of the energy distribution.


\section{\label{sec:conclusions} Conclusions}


In this paper the main attention is paid to a possible non-Hermiticity
of the Yukawa  interaction between the Higgs scalar field with
fermions. A model for non-Hermitian interaction is proposed and
approximation of  one fermion generation is considered. The
corresponding Lagrangian is obtained,  and for the free fermion the
modified Dirac equation, which contains the ``mass'' term  in the
form $m_1+ m_2 \gamma_5$, is studied.
The symmetry of the Lagrangian with respect to the discrete
$\mathcal{P}$, $\mathcal{C}$ and $\mathcal{T}$ transformations is
addressed, in particular, for real parameters $m_1$ and $m_2$ the
Lagrangian appears to be $\mathcal{P}$-odd,
$\mathcal{C}$-even, $\mathcal{T}$-even, $\mathcal{CPT}$-odd and non-Hermitian.

We discuss the decay of the Higgs boson to the polarized fermion $f$
and  antifermion $\bar{f}$, and calculated the decay rate and
polarization characteristics of $f, \, \bar{f}$. The interaction
vertex $h f \bar{f}$ is parametrized in terms of the two
couplings $a_f$ ($\mathcal{CP}$-even term) and $b_f$
($\mathcal{CP}$-odd term) in such a way that for a general case of
complex  $a_f$ and $b_f$ the interaction is non-Hermitian. This
non-Hermiticity gives rise to polarization of fermion and
antifermion along the direction of their movement. The magnitude
of the longitudinal polarization is determined by the factor
$\propto {\rm Im}(a_f b_f^*)$.

In connection with violation of the $\mathcal{CPT}$ symmetry and non-Hermiticity in the present model, we note that most frequently the
$\mathcal{CPT}$ symmetry is tested via measurement of the differences between the masses of particle and its antiparticle and some other their characteristics (see, for example,~\cite{PDG:2014}).
These experiments are based on the $\mathcal{CPT}$ theorem which
is a consequence of Lorentz invariance, locality, connection between spin and statistics, and a Hermitian Hamiltonian~\cite{Streater:1964}.
Nevertheless, even if the masses of particle and antiparticle are equal, the $\mathcal{CPT}$ invariance can be violated in scattering and other physical processes~\cite{Greenberg:2002}. It is also proved~\cite{Greenberg:2002} that the $\mathcal{CPT}$ violation leads to violation of the Lorentz invariance.

Unlike the case of the particle-antiparticle mass difference, the longitudinal polarization of the fermion in the decay $h \to f \bar{f}$ is an example of the $\mathcal{CPT}$-violating observable in Lorentz invariant but non-Hermitian model. Another such observable is the circular polarization of the photon in the Higgs-boson decays $h \to \gamma \gamma$ and $h \to \gamma Z$ \cite{Korchin:2013a} (other examples and detailed discussion are given in Ref.~\cite{Okun:2002}). In general, non-Hermitian Lagrangian (or Hamiltonian) leads to violation of the unitarity of the ${\mathcal S}$-matrix, however measurement of the longitudinal polarization of the fermion can be easier task than direct tests of the unitarity violation.

In order to search for the fermion longitudinal polarization we
considered the Higgs boson decay to the $\tau^- \tau^+$ leptons
with their subsequent decay into the leptonic channels, i.e. the
process $h \to \tau^- \tau^+ \to \mu^- {\bar \nu}_\mu \nu_\tau \,
\mu^+ \nu_\mu {\bar \nu}_\tau$. For this decay the fully
differential decay width and the distribution over the energies of
the muons $\mu^-$ and $\mu^+$ are analytically derived. Then an
observable is proposed, called the asymmetry, which is nonzero
for a non-Hermitian $h \tau^- \tau^+$ interaction.

This asymmetry has the form of a product of non-Hermiticity factor
$\propto \! \! \! \! {\rm Im}(a_f b_f^*)$ and function
$\Delta (\varepsilon_1,\varepsilon_1^\prime;\varepsilon_2,\varepsilon_2^\prime)$,
which depends on the area of energies of $\mu^-$ and $\mu^+$. We calculated this function for various configurations of muon energies and selected
optimal conditions for studying this observable.  Other observables
proportional to ${\rm Im}(a_f b_f^*)$ are also studied and calculated.

We hope that the study of the asymmetries in the decay $h \to \tau^- \tau^+ \to \mu^- {\bar \nu}_\mu \nu_\tau \, \mu^+ \nu_\mu {\bar \nu}_\tau$, considered in the present paper, will be useful for the test of Hermiticity of the Yukawa interaction.


\section*{ACKNOWLEDGMENTS}
This research was partially supported by National
Academy of Sciences of Ukraine (project TsO-1-4/2016) and Ministry of Education and Science of Ukraine (project no. 0115U000473).

\vspace*{\fill}


\appendix


\section{ DEFINITION OF FUNCTIONS $f(x_1,x_2)$ AND $g(x_1,x_2)$ }
\label{sec:Appendix}

The functions $f(x_1,x_2)$ and $g(x_1,x_2)$ which enter the energy distribution in Eq.~(\ref{eq:2007}) have the form
\begin{widetext}
\begin{eqnarray}
&&f(x_1,x_2)=a(x_1)a^2(x_2)/3-2(1+\beta)(1-y)a^2(x_2)/3-(1+\beta)(2+y)a(x_1)a(x_2)/4+
(1+\beta)^2(1-y^2)a(x_2)\nonumber
\\&&-(1+\beta)^3y(1-y)^2+(1-\beta)^{-1}\Bigl(x_1x_2a(x_1)a(x_2)-2(1+\beta)(1-y)x_1x_2a(x_2)+(1+\beta)^2(1-y)^2x_1x_2\nonumber
\\&&-a(x_1)b(x_1)\frac{a(x_2)b(x_2)}{36\beta^2}+(1+\beta)(1-y)a(x_2)b(x_2)
\frac{c(x_1)}{12\beta^2}-(1+\beta)^2(1-y)^2\frac{c(x_1)c(x_2)}{16\beta^2}\Bigr),
\label{eq:A001}
\end{eqnarray}
\begin{eqnarray}
&&g(x_1,x_2)=(2x_1-1)a(x_1)a^2(x_2)/3+2(1+\beta)(1-y)(1-x_1)a^2(x_2)/3-(1+\beta)^2(1-y)
(2-x_1-2x_2\nonumber
\\&&+y(2x_2-x_1))a(x_2)/2+
(1+\beta)^3(1-y)^3x_1-(1+\beta)(1+y)x_1a(x_1)a(x_2)/2.
\label{eq:A002}
\end{eqnarray}
\end{widetext}
Here
\begin{eqnarray}
a(x)&\equiv&(1+\beta)(1+y)-2\,z(x)-2\, \frac{x-z(x)}{1-\beta},
\label{eq:A003}\\
b(x)&\equiv&2x+4\beta\, z(x)-(1-\beta^2)(1+y),\label{eq:A004}\\
c(x)&\equiv&2x+2\beta\, z(x)-(1-\beta^2)(1+y),\label{eq:A005}
\end{eqnarray}
and $z(x)\equiv\sqrt{x^2-y(1-\beta^2)}$.


\end{document}